\documentclass[submit]{smj}

\usepackage{graphicx,setspace,lscape,longtable}
\usepackage{mathrsfs,amsthm}
\usepackage{natbib,epsfig,pdfpages}
\usepackage{rotating}
\usepackage{float}
\usepackage{natbib}
\usepackage{bm}
\usepackage[normalem]{ulem}
\usepackage{CJK}
\usepackage{subfigure}
\usepackage{array,booktabs}
\usepackage[titletoc]{appendix}
\usepackage{algorithm}
\usepackage{algpseudocode}

\bibpunct{(}{)}{;}{a}{,}{,}

\def\beq{\begin{equation}}
\def\eeq{\end{equation}}
\def\beqr{\begin{eqnarray}}
\def\eeqr{\end{eqnarray}}
\def\beqrs{\begin{eqnarray*}}
\def\eeqrs{\end{eqnarray*}}
\def\bc{\begin{center}}
\def\ec{\end{center}}

\Author{Tingting Huang\Affil{1},
        Gilbert Saporta\Affil{2},
        Huiwen Wang\Affil{1,3},
        and Shanshan Wang\Affil{1,4}
}
\AuthorRunning{Shanshan Wang \textrm{et al.}}

\Affiliations{

\item School of Economics and Management,
      Beihang University,
      Beijing,
      China
\item CEDRIC CNAM, 292 rue St Martin, 75141 Paris Cedex 03, France
\item Beijing Advanced Innovation Center for Big Data and Brain Computing,
      Beihang University,
      Beijing,
      China
\item Beijing Key Laboratory of Emergence Support Simulation Technologies for City Operations,
      Beijing,
      China

}   

\CorrAddress{Shanshan Wang,
             School of Economics and Management,
             Beihang University,
             Xueyuan Road No.37,
             Haidian District,
             Beijing,
             China}
\CorrEmail{sswang@buaa.edu.cn}
\CorrPhone{(+86)\;010\;82339337}
\CorrFax{(+86)\;010\;82328037}

\Title{Spatial Functional Linear Model and Its Estimation Method}
\TitleRunning{Spatial Functional Linear Model}

\Abstract{
The classical functional linear regression model (FLM) and its extensions, which are based on the assumption that all individuals are mutually independent, have been well studied and are used by many researchers. This independence assumption is sometimes violated in practice, especially when data with a network structure are collected in scientific disciplines including marketing, sociology and spatial economics. However, relatively few studies have examined the applications of FLM to data with network structures. We propose a novel spatial functional linear model (SFLM), that incorporates a spatial autoregressive parameter and a spatial weight matrix into FLM to accommodate spatial dependencies among individuals. The proposed model is relatively flexible as it takes advantage of FLM in handling high-dimensional covariates and spatial autoregressive (SAR) model in capturing network dependencies. We develop an estimation method based on functional principal component analysis (FPCA) and maximum likelihood estimation. Simulation studies show that our method performs as well as the FPCA-based method used with FLM when no network structure is present, and outperforms the latter when network structure is present. A real weather data is also employed to demonstrate the utility of the SFLM.
}


\begin{document}

\maketitle

\section{INTRODUCTION}\label{sec1}

\noindent With the rapid development of electronic technology, huge amounts of data can be collected and stored cheaply. In particular, some types of data that are recorded at high frequencies can be regarded as almost continuously observed. Examples include trajectory data, weather data and stock data. These types of data are called functional data, which are inherently infinite-dimensional and are rich in information. Hence, functional data analysis (FDA) has been applied in many subject areas, such as biology, the medical sciences, meteorology, econometrics, finance, chemometrics and geophysics. For examples, see \cite{RamsaySilverman-506} and \cite{horvath2012inference}.

Functional regression is one of the most important tools in FDA. Classical functional linear regression model (FLM) represents the simplest form of functional regression (an overview is given in \cite{paganoni2017functional}), and receives much attention in FDA. Studies related to FLM include \cite{TT2006Prediction,Hall2007Methodology,ferraty:hal-00966065,Zhou2013FUNCTIONAL,brockhaus2015functional}, and reference therein. FLM has become increasingly popular because they can be applied to problems that are difficult to address in the framework of scalar and vector observations.

Generally, let $Y$ be a scalar response variable, and let $X(t)$ be a second-order stochastic process on a compact interval $\Gamma$. Moreover, we assume that $X(t)$ is square integrable with zero mean
and $E(\int_{\Gamma}X^{2}(t)dt)<\infty$. Formally, a FLM can be written as
\beq\label{flm}
Y=\alpha+\int_{\Gamma}X(t)\beta(t)dt+\epsilon,
\eeq
where $\alpha$ is the intercept, $\beta(t)$ is the unknown slope function, and $\epsilon$ is the random error term. This term is independent of $X(t)$ and has zero mean and finite variance.

During the past few years, extensions of the FLM (\ref{flm}) have been studied to address specific problems. For example, \cite{RSSB:RSSB342} proposed functional logistic regression and functional censored regression to address cases in which the responses of FLM is binary and right censoring, respectively. Subsequently, \cite{Escabias2004Principal} presented some alternative methods for estimating the parameter function in functional logistic regression model based on principal components. \cite{Aneiros2006Semi} constructed a semi-functional partial linear model that combines the advantages of semi-linear model and nonparametric statistics for functional data. \cite{ferraty:hal-00966065} generalized the FLM to functional projection pursuit regression, which allows for more interpretability. \cite{Liu2017Estimating} presented a functional linear mixed model to investigate the scalar and functional covariate effects of both individuals and population.

All of the methods mentioned above assume that all the individuals are mutually independent. However, given the rapid advances in information technology, relation information among individuals can easily be collected. Network-structured data, which represent one common form of relation information, are becoming increasingly available. In such data, the realizations of the dependent variable are correlated with each other. If we use FLM or its aforementioned variations of FLM to model this kind of data, the information contained in the data may not be fully exploited; moreover, the inferences obtained when network effects are ignored may be misleading. These observations motivate us to develop a novel statistical model for application to functional data with network structures.

\begin{figure}[!h]
  \begin{center}
  \subfigure{
  \label{fig1-1}
    \resizebox{6cm}{5cm}{\includegraphics{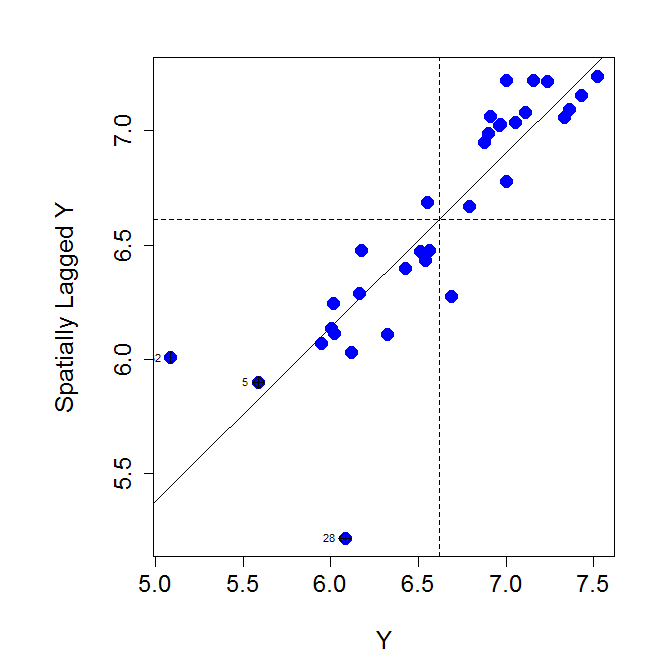}}}
    \subfigure{
    \label{fig1-2}
    \resizebox{6cm}{5cm}{\includegraphics{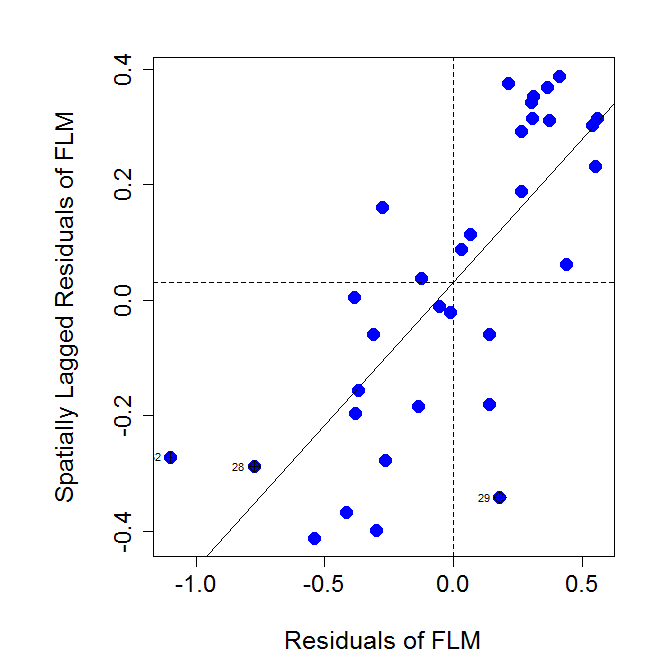}}}
  \end{center}
  \vspace{-0.5cm}
  \caption{Moran's I scatterplot of $Y$ (left) and the residuals of the FLM (right), respectively} \label{Fig.1}
\end{figure}

To better illustrate our motivation for carrying out this study and its significance, we show a motivating example. We collected weather data from the China Meteorological Yearbook covering the period between $2005$ and $2007$. These data record monthly temperatures and precipitation in $34$ major cities in China. Our aim is to investigate the effect of temperature on precipitation over these three years. The scalar response is the logarithm of the mean annual precipitation, and the functional covariate is the mean monthly temperature. In a preliminary analysis, we employ the Moran's I statistic to test whether spatial autocorrelation exists among the responses. Unfortunately, the value of the Moran's I statistic is $0.7$, and the P value is smaller than $0.001$, which indicates that there is a significant correlation among these responses. After  we apply the FLM directly to the weather data, the spatial dependence among the residuals of the FLM persists; the Moran I statistic is $0.5$, and its P value is smaller than $0.001$. We also display the Moran's I scatterplot of the response $Y$ and the residuals of the FLM in Figure \ref{Fig.1}.

\vspace{10pt}

Figure \ref{Fig.1} shows that the responses and the residuals of FLM display a significance spatial autocorrelation. Thus, FLM may not be appropriate for such data.  This result motivates the incorporation of spatial correlation in analyses of spatially correlated functional data. A detailed analysis of this weather data can be found in Section \ref{sec5}.

Fortunately, when the predictor is scalar, instead of functional, spatial autoregressive (SAR) model is frequently used model to accommodate the dependencies introduced by network structures. SAR model has been applied to social interactions, strategic interactions and geostatistics. Studies that apply SAR model include \cite{Case1991Spatial}, \cite{Topa2001Social}, and \cite{Olubusoye2016MODELLING}, among others. Moreover, estimation methods for SAR model are described in \cite{Ord1975Estimation}, \cite{Lee2004Asymptotic}, \cite{Kelejian1999A}, \cite{Lee2007GMM} and \cite{Lesage2009Introduction}. In a SAR model, a spatial weight matrix is employed to denote adjacent relations among the observations, and an unknown spatial autoregressive parameter is used to reflect the strength of neighbouring effects. By borrowing concepts from SAR model, it is straightforward to incorporate the network structures into FLM model using a spatial autocorrelation parameter and a weight matrix. The proposed model is here named the spatial functional linear model (SFLM).

The new SFLM displays powerful capabilities in addressing the functional covariate in the model, together with the spatial dependencies between the outcomes of adjacent units. To obtain estimates of the unknown parameters, we develop an easily implemented estimation method that employs the FPCA method to handle functional data; moreover, maximum likelihood estimator for SAR model (\cite{Ord1975Estimation}, \cite{Lee2004Asymptotic}) is used to handle the spatial parameters. The simulation results show that our estimation method performs well. In particular, the new method behaves as well as the FPCA-based method used with FLM  when the spatial autocorrelation parameter equals zero, in which case the SFLM degenerates into an FLM; moreover, the new method displays better performance than the latter when network dependencies are present. A real dataset is employed to illustrate the practical utility of the proposed model.

We also note that \cite{Pinedar2016Functional} proposed a model that combines a spatial error model (SEM) with an FLM to accommodate network structure in functional data. The most important difference between our model and his is that the spatial dependencies in our model are contained in the response, whereas they are contained in the disturbances in the existing model. Pinedar also used maximum likelihood estimation method to estimate the parameters of his model; however, he did not provide a practical implementation method. Moreover, our methods, which were developed for use with SFLM, can be easily extended to his functional SAR model and other spatial regression models. Note that our model handles scalar responses. This feature differs strongly from those of \cite{Aguilera2016Prediction} and  \cite{Giraldo2017Spatial}, which consider functional responses in spatial functional data.


\textcolor[rgb]{1.00,0.00,0.00}{ }

We organise this article as follows. In Section \ref{sec2}, we formulate the new model. The proposed estimation method is constructed based on FPCA and the MLE in Section \ref{sec3}. The finite-sample performance of the proposed estimators is evaluated through simulation studies and is compared with the finite-sample performance of the FLM in Section \ref{sec4}. Finally, in Section \ref{sec5}, we use real data to document the usefulness of this methodology. We conclude the article with a discussion in Section \ref{sec6}.

\section{MODEL SPECIFICATION}\label{sec2}

\noindent Following \cite{Qu2015Estimating}, we consider the spatial processes located on an unevenly spaced lattice $D\subseteq R^{d}$, $d\geq1$. We observe $\{(x_i (t) ,y_i )\}^{n}_{i=1}$ from $n$ spatial units on $D$. Here, $x_i (t)$ represents independent and identically distributed (i.i.d.) samples from $X(t)$, where $X(t)$ is a square integrable second-order stochastic process defined on a compact set $\Gamma$ with $E(X(t))=0$ and $E(\int_{\Gamma}X^2(t)dt)<\infty$. Without loss of generality, we presume $\Gamma$ is the unit, i.e. $t\in[0,1]$.

We denote $\bm{x}(t)=\big(x_1(t),x_2 (t),\cdots,x_n (t)\big)^\prime$ and $\bm{y}=(y_1,y_2,\cdots,y_n)^\prime$. We then formulate the SFLM as
\beq\label{sflm}
\bm{y}=\alpha\bm{\tau}_n+\rho \bm{Wy}+\int_{0}^{1}\bm{x}(t)\beta(t)dt+\bm{\epsilon},
\eeq
where $\alpha \bm{\tau}_n$ is the intercept. Here $\bm{\tau}_n$ represents an n-dimensional vector of ones; $\alpha$ denotes a scalar parameter; $\rho$ is the unknown spatial autocorrelation parameter that takes values within the range of $[0, 1)$; $\bm{W}=(w_{ii^{\prime}})_{n\times n}$ is a pre-specified spatial weight matrix, in which $w_{ii^\prime}$ represents the weight between units $i$ and $i^\prime$; $\beta(t)$ is a square integrable coefficient function defined on $[0,1]$; and $\epsilon$ is the noise term, which is independent of $\bm{\bm{x(t)}}$. We assume that $\bm{\epsilon}$ follows a multivariate normal distribution with zero mean and a constant diagonal covariance matrix $\sigma^2 \bm{I}_n$, where $\bm{I}_n$ is a $n\times n$ identity matrix.

In the model presented in (\ref{sflm}), the spatial weight matrix $\bm{W}$ is exogenous. This matrix is constructed according to the distances between the units on $D$ in different contexts. For the case of geographic locations, $\bm{W}$ is formed based on adjacency relations or the nearest $k$ neighbours, as measured in terms of the Euclidean distance or another appropriate metric. In social networks, $W$ is established on the basis of friend relationships among individuals. In other practical settings, such as economics, $\bm{W}$ can also be built using economic factors such as GDP.  In general, the spatial matrix $\bm{W}$ is standardized to be a row-normalized matrix; in this matrix, the summation of the row elements is unity, and the entries on the diagonal are zeros.

In addition, $\rho$ is a scale parameter that reflects the impact of the neighbours. Greater values of $\rho$ indicate that $y_i$ is more strongly affected by its neighbours. The SFLM clearly reduces to the classical FLM when $\rho=0$ and the SAR model when $\bm{\bm{x(t)}}$ is a constant covariate that does not depend on $t$. Thus, the proposed SFLM is a more flexible model, given its incorporation of both functional covariates and network structures.

To provide better insight into the new model, we divide right-hand side of equation (2) into two parts. The first part is $\rho \bm{Wy}$, which represents the neighbour effects. In real cases, such effects may arise due to competition, spillover or shared sources. The second part, $\int_{0}^{1}\bm{\bm{x(t)}}\beta(t)dt$, indicates the effects of exogenous functional variables. The outcome of unit $i$ is affected by the outcomes of its neighbours' $i^\prime$ ($i^\prime \neq i$) as well as its individual-specific covariate $x_i (t)$.

Morover, we can reformulate equation (\ref{sflm}) as the following equivalent expression,
\beq\label{sflm-1}
\bm{y}=(\bm{I}_{n}-\rho \bm{W})^{-1}\alpha \bm{\tau}_n+(\bm{I}_{n}-\rho \bm{W})^{-1}\int_{0}^{1}\bm{\bm{x}(t)}\beta(t)dt+(\bm{I}_{n}-\rho \bm{W})^{-1}\bm{\epsilon},
\eeq
which shows how $y$ is generated. $y_i$ is clearly influenced by its neighbours' covariates $x_{i^\prime }(t)$s, as is $i^\prime\neq i$. The error terms in equation (\ref{sflm-1}) are no longer independent. Given that this dependency violates the Gauss-Markov assumption, the ordinary least-squares (OLS) estimator is not adequate to estimate the parameters in (\ref{sflm}).

\section{ESTIMATION METHOD}\label{sec3}

\noindent We estimate the intercept $\alpha$, the spatial autocorrelation parameter $\rho$, the slope function $\beta(t)$, and the variance of the error term $\sigma^2$ in the SFLM (\ref{sflm}) using the maximum likelihood approach combined with a basis expansion in terms of the FPCA, as shown follows.

\subsection{Basis expansion based on FPCA}\label{sec3-1}

\noindent Note that, before estimating the parameters of our model, data representation methods, such as smoothing and interpolation, should be used to convert discretely recorded data $x_i (t_i)$ to a curve $x_i (t)$.

Let $K(s,t)$ denote the covariance function for $X(t)$, i.e. $K(s,t)=\mbox{Cov}(X(t),X(s))$. By Mercer's theorem, the spectral decomposition of $K(s,t)$ is then $$K(s,t)=\sum_{j=1}^{\infty}k_j \varphi_j (s) \varphi_j(t)$$, where $k_1>k_2>\cdots>0$ are eigenvalues and $\{\varphi_j(t)\}_{j=1}^{\infty}$ are the corresponding orthogonal eigenfunctions. According to the Karhunen-Lo\`{e}ve expansion, 
$X(t)$ can be expanded as $$X(t)=\sum_{j=1}^{\infty}a_j \varphi_j (t),$$ where the $a_j$s are uncorrelated random variables with mean zero and variance $E(a_j^{2})=k_{j}$ with $a_j=\int_{0}^{1} X(t)\varphi_j(t)dt$. 

For an observation $\{y_i,x_i(t)\}_{i=1}^n$, the empirical version of $K(s,t)$ is $$\hat{K}(s,t)=\frac{1}{n} \sum_{i=1}^{n}x_i (s) x_i (t)- \bar{x}(s) \bar{x}(t),$$ where $\bar{x}(t)=\frac{1}{n} \sum_{i=1}^{n}x_i (t)$. Moreover, it can be shown that
$$\hat K(s,t)=\sum_{j=1}^{n}\hat k_j \hat \varphi_j (s) \hat\varphi_j(t),$$
where $\hat k_j$ and $\hat\varphi_j(t)$ are the estimators of $\varphi_j(t)$ and $k_j$, respectively. For the $i$th observation $x_i(t)$, the estimator of $a_{ij}$ is then $\hat a_{ij}=\int_0^1 x_i(t)\hat\varphi_j(t)$, and
$x_i (t)$ can be written as $x_i (t)=\sum_{j=1}^{\infty}\hat{a}_{ij} \hat{\varphi}_j (t)$.

Similarly, based on the estimated orthonormal functional basis $\{\hat\varphi_j(t)\}_{j=1}^{\infty}$, $\beta(t)$ has the expression $\beta(t)=\sum_{j=1}^{\infty}b_j \hat\varphi_j (t)$, with $b_j=\int_{0}^{1}\beta(t) \hat\varphi_j(t)dt$.
Therefore, SFLM has the equivalent expression
\beq\label{sflm-2}
\bm{y}=\alpha \bm{\tau}_n+\rho \bm{Wy}+\sum_{j=1}^{n}\bm{\hat{a}_{j}} b_j +\bm{\epsilon},
\eeq
where $\bm{\hat{a}_j}=(\hat{a}_{1j}, \hat{a}_{2j}, \cdots, \hat{a}_{nj})^\prime$.

In reality, the first few principal components (PCs), as ranked by their eigenvalues, often provide an adequate approximation of $x_i (t)$. Here, we choose the first $m$ PCs and ignore the truncation error. The truncated SFLM of (\ref{sflm-2}) takes the form of
\beq\label{appro-sflm}
\bm{y}\approx\alpha \bm{\tau}_n+\rho \bm{Wy}+\sum_{j=1}^{m}\bm{\hat{a}_{j}} b_j+\bm{\epsilon}.
\eeq

\subsection{Maximum Likelihood Estimation for the Truncated SFLM}\label{sec3-2}

\noindent Here we focus on (\ref{appro-sflm}) and adopt the popular MLE approach to estimate the unknown parameters.

Defining $\bm{A}= ( \hat{a}_{ij})_{n\times m}$, $\bm{b}=(b_1, b_2, \cdots, b_m)^\prime$, $\bm{Z}=(\bm{\tau}_n,\bm{A})$ and $\bm{\delta}=(\alpha,\bm{b}^{\prime})^{\prime}$, we can write the truncated equation (5) as
\beq\label{appro-sflm-1}
\bm{y}\approx\rho \bm{Wy}+\bm{Z\delta}+\bm{\epsilon}.
\eeq
Based on the assumption that the error term in (\ref{sflm}) follows a multivariate normal distribution, the log-likelihood function for $\bm{y}$ is
\beq\label{log-lik}
\ln L(\rho, \bm{\delta}, \sigma^2)=-\frac{n}{2}\ln(2\pi\sigma^2)+\ln|\bm{I}_n-\rho \bm{W}|-\frac{\bm{e}^\prime \bm{e}}{2\sigma ^2},
\eeq
where $\bm{e}=\bm{y}- \rho \bm{Wy}- \bm{Z\delta}$. Given $\rho$, the maximum likelihood estimate of $\bm{\delta}$ and $\sigma^2$ is
\begin{eqnarray}
  \bm{\hat{\delta}}(\rho) &=& (\bm{Z}^\prime \bm{Z})^{-1}\bm{Z}^\prime(\bm{\bm{I}_n}-\rho \bm{W})\bm{y},\label{deltahat} \\
  \hat{\sigma}^2 (\rho) &=& \frac{1}{n}(\bm{y}-\rho \bm{Wy}-\bm{Z}\bm{\hat{\delta}})^{\prime}(\bm{y}-\rho \bm{Wy}-\bm{Z}\bm{\hat{\delta}}(\rho)).\label{sigmahat}
\end{eqnarray}
Substituting $\bm{\hat{\delta}}(\rho)$ and $\hat{\sigma}^2 (\rho)$ into (\ref{log-lik}) and dropping the constant term yields the following profile log-likelihood function
\beq\label{log-lik-1}
\ln L(\rho)=-\frac{n}{2}\ln(\hat{\sigma}^2(\rho))+\ln|\bm{I}_n-\rho \bm{W}|,
\eeq

The optimization of the maximum log-likelihood function (\ref{log-lik-1}) then reduces to a univariate optimization problem.
That is, the estimator of $\rho$ can be obtained via the maximization of (\ref{log-lik-1}), i.e. \begin{equation}\label{rhohat}
\hat\rho=\mbox{arg}\max_\rho\ln L(\rho).\end{equation} 
Replacing $\rho$ with $\hat\rho$ in (\ref{deltahat}) and (\ref{sigmahat}) yields the estimators of $\bm{\delta}$ and $\sigma^2$, respectively. The estimator $\hat{\beta}(t)$ for $\beta(t)$  can be reconstructed by
\beq\label{betat}
\hat{\beta}(t)=\sum_{j=1}^{m}\hat{b}_j\hat{\varphi}_j(t).
\eeq 
where $\hat{b}_j$ is obtained from $\bm{\hat{\delta}}$ in (\ref{deltahat}).

In summary, the estimators $\hat{\rho}$, $\widehat{\alpha}$ and $\widehat{\sigma}^2$ of SFLM are obtained from the MLE of the truncated SFLM; on the other hand, the estimator $\hat{\beta}(t)$ of the SFLM is constructed using the functional principal component basis presented in Section 3.1 and the coefficient estimator given in Section 3.2. For convenience, we here summarize the main steps of the estimation procedure in Algorithm 1.

\vspace{3mm}
\begin{algorithm}[!h]
  \caption{Main steps of the estimation procedure}
  \begin{algorithmic}[1]
    \State Represent the functional predictor and slope function using the functional principal component basis. The estimation procedure is then simplified, as the remaining process is similar to the problem of estimating a SAR model. In this step,  after an appropriate truncation parameter is given, the SFLM approximates a SAR model whose covariates are the principal component scores of $x_{i}(t)$s, as shown in (\ref{appro-sflm-1}).
    \State Determine the estimators of the unknown parameters of the truncated SFLM obtained in Step 1. The maximum likelihood estimation method is used to obtain the spatial autocorrelation parameter $\rho$ in (\ref{rhohat}), the estimators of the coefficients $b$, the intercept $\alpha$ and the variance of the error term $\sigma^2$, as shown in (\ref{deltahat})-(\ref{sigmahat}).
    \State Determine the estimator of $\beta(t)$ in the SFLM. The slope function is constructed using the FPC basis mentioned in Step 1 and the coefficients estimated in Step 2, as shown in (\ref{betat}). The other estimators are obtained directly from Step 2.
  \end{algorithmic}
\end{algorithm}


\subsection{Choosing the truncation parameter for the SFLM}\label{sec3-3}
There are two ways to determine the truncation parameter. The first method is the percentage of variance explained (PVE) for predictors, which uses eigenvalues to choose the number of PCs. The second method is derived from a Markov chain Monte Carlo model composition methodology labeled MC$^3$.

To compare the new SFLM with the FLM in numerical experiments, we use the first method to determine the values of this parameter. In this case, if we choose the PVE to be $80\%$, the truncation parameter $m$ is subject to $\underset{l}{\mbox{min}}~\{(\sum_{j=1}^{l}\hat{k}_{j})/(\sum_{j=1}^{\infty}\hat{k}_j)\geq80\%\}$.

The second method specifies the covariates in (\ref{appro-sflm-1}) according to the posterior model probability. We adopt this technique to handle the weather data in Section \ref{sec5}. In this Bayesian methodology, prior distributions are assigned to the parameters in the truncated SFLM. Specifically, $\sigma^2$ follows an inverse gamma distribution, $\pi(\sigma^2)\sim IG(a,b)$; $\bm{b}^{\prime}$ follows a multivariate normal distribution conditional on $\sigma^2$, i.e. $\pi(\bm{b}^{\prime}| \sigma^2)\sim N[\bm{b}^{\prime}_0,\sigma^2(g \bm{A}^{\prime} \bm{A})^{-1}]$; and $\rho$ follows a Beta prior distribution, $\pi(\rho)\sim \frac{1}{Beta(d,d)} \frac{(1+\rho)^{d-1}(1-\rho)^{d-1}}{2^{2d-1}}$. We set $a=0,b=0,g=\frac{1}{n},d=1.01$ for use with the weather data. Note that $\bm{A}$ must be scaled during preprocessing. Details of this procedure are given in \cite{Lesage2007Bayesian}.

\section{SIMULATION STUDY}\label{sec4}

\noindent We conduct several simulation studies to evaluate the finite-sample performance of the proposed estimators of $\rho$ and $\beta(t)$. All of the computations were carried out in the R environment, and we use existing functions in the R packages `spdep', `fda' and `fda.usc' to implement the proposed procedure.

Because spatial networks are of interest in this study, we compare the proposed SFLM with the existing FLM in terms of the behaviour of the estimators of $\beta(t)$. In particular, the SFLM is estimated using the proposed method, whereas the FLM employs an FPCA-based estimation approach. To make these two estimation methods comparable, we set the truncation parameter of FPCA to be identical to the PVE, which equals $70\%$.
Different degrees of spatial effects are considered; the spatial parameter $\rho$ is set to $0$, $0.5$, and $0.8$. Note that,  when $\rho=0$, the SFLM reduces to the FLM.

As for the spatial scenario, we adopt the rook matrix by randomly apportioning $n$ agents on a regular square grid of cells; each agent is located on a cell. In this context, if the grid has $R$ rows and $T$ columns, then the sample size $n=R\times T$. Units that share an edge are neighbours. This definition ensures the units in the inner field of the grid have four neighbours, the units in the corners have two neighbours, and the units along the boarders have three neighbours. Therefore, the spatial matrix is an adjacent matrix with each entry $w_{ii^\prime}=1$ if units $i$ and $i^\prime$ are neighbours and $w_{ii^\prime}=0$ otherwise. We set $n=\{10\times 30, 20\times 25, 30\times30\}$ in the simulation. The spatial weight matrix is row-normalized in all cases.

For the functional part of equation (2), we emply the same form as the functions in the FLM by \cite{Hall2007Methodology}. Specifically, we generate the simulation data $\bm{y}=(y_1, y_2, \cdots, y_n)^{\prime}$ using
$$\bm{y}=(\bm{I}_n-\rho \bm{W})^{-1}\Big(\int_{0}^{1}\bm{x}(t)\beta(t)dt+0.5\bm{\epsilon}\Big),~~~~  \epsilon_i\sim N[0, 1],$$
where the spatial parameter $\rho$ is assigned values of $0$, $0.5$, and $0.8$, respectively. The functional predictor $\bm{x}(t)=(x_1(t), x_2(t), \cdots, x_n(t))^{\prime}$ is produced with values of $x_i(t)$ that are independently generated as
$$x_i(t)=\sum_{j=1}^{50}a_{j}Z_{j}\varphi_j(t),$$
where $a_j=(-1)^{j+1}j^{-\gamma/2}$ with $\gamma=1.1$ and $2$, respectively; $Z_j\sim U[-\sqrt{3}, \sqrt{3}]$ and $\varphi_j(t)=\sqrt{2}\cos(j\pi t)$. Similarly, the coefficient function $\beta(t)$ is generated according to
$$\beta(t)=\sum_{j=1}^{50}b_{j}\varphi_j(t),$$
where $b_1=0.3$ and  $b_j=4(-1)^{j+1}j^{-2}, j\geq2$.





Note that the eigenvalues of the covariance function $\hat{K}(u,v)$ play a vital role in determining the estimation accuracy of $\beta(t)$. We consider two cases. In case 1, $\gamma=1.1$; thus the 
eigenvalues are well spaced and the slope function can be accurately estimated. In case 2, $\gamma=2$, and the 
closely spaced eigenvalues can cause the estimator $\hat{\beta}(t)$ to display poor performance.

The experiment is repeated $500$ times in each setting. We assess the performance of the estimator $\hat{\rho}$ in terms of the mean bias and standard derivation; on the other hand, we evaluate the performance of the estimator of $\beta(t)$ in terms of the mean square error (MSE) evaluated at $100$ equispaced points on $[0,1]$, $\{t_i\}_{i=1}^{100}$; i.e.,
$$\mbox{MSE}=\frac{1}{100}\sum_{i=1}^{100} \big(\hat\beta(t_i)-\beta(t_i)\big)^2,$$
where $\hat\beta(t_i)$ is the estimator of $\beta(t_i)$ evaluated at $t_i$, which is obtained via the SFLM or the FLM. We summarize these results in Table \ref{Tab 1}.

\begin{table}[!h]
  \caption
  {The empirical average biases and standard deviations (in brackets) of the estimator of $\rho$, denoted $\mbox{bias(sd)}$, and the empirical average MSE and its standard deviations (in brackets) of $\beta(t)$ obtained via SFLM and FLM, denoted $\mbox{MSE}_1\mbox{(sd)}$ and $\mbox{MSE}_2\mbox{(sd)}$, respectively.}
  \label{Tab 1}
  \setlength\tabcolsep{6pt}
  \renewcommand{\arraystretch}{1.2}
  \begin{center}
    \begin{tabular}{ccccccccc}
    \hline
 &&\multicolumn{3}{c}{$\gamma=1.1$}&&\multicolumn{3}{c}{$\gamma=2$}\\
 \cline{3-5}\cline{7-9}
$\rho$ & $n$ & $\mbox{bias(sd)}$ & $\mbox{MSE}_1\mbox{(sd)}$ & $\mbox{MSE}_2\mbox{(sd)}$ & & $\mbox{bias(sd)}$ & $\mbox{MSE}_1\mbox{(sd)}$ & $\mbox{MSE}_2\mbox{(sd)}$\\
\hline
 0 & 300 & $\underset{(0.0495)}{-0.0051}$ & $\underset{(0.0072)}{0.0203}$ & $\underset{(0.0072)}{0.0203}$ & & $\underset{(0.0628)}{-0.0086}$ & $\underset{(0.0239)}{0.1171}$ & $\underset{(0.0239)}{0.1171}$ \\
 &500 &$\underset{(0.0428)}{-0.0003} $& $\underset{(0.0030)}{0.0087} $&$\underset{(0.0029)}{ 0.0087} $& &$\underset{(0.0459)}{ -0.0020} $& $\underset{(0.0109)}{0.0691} $& $\underset{(0.0109)}{0.0691}$\\
 & 900 & $\underset{(0.0294)}{-0.0024}$ &$\underset{(0.0011)}{ 0.0034}$ &$\underset{(0.0011)}{ 0.0034 }$& &$\underset{(0.0369)}{ 0.0006}$ & $\underset{(0.0044)}{0.0378}$ & $\underset{(0.0044)}{0.0378}$\\
 0.5&300&$\underset{(0.0457)}{-0.0062}$ & $\underset{(0.0071)}{0.0201}$ & $\underset{(0.0101)}{0.0267}$ && $\underset{(0.0524)}{-0.0068}$ & $\underset{(0.0226)}{0.1173}$ & $\underset{(0.0227)}{0.1198}$ \\
 &500 & $\underset{(0.0343)}{-0.0016}$ & $\underset{(0.0027)}{0.0085}$ & $\underset{(0.0036)}{0.0114}$ && $\underset{(0.0400)}{-0.0037}$ & $\underset{(0.0101)}{0.0689}$ & $\underset{(0.0102)}{0.0702}$ \\
 &900& $\underset{(0.0241)}{-0.0034}$ & $\underset{(0.0010)}{0.0034}$ & $\underset{(0.0013)}{0.0047}$ && $\underset{(0.0292)}{-0.0046}$ & $\underset{(0.0045)}{0.0380}$ & $\underset{(0.0046)}{0.0386}$ \\
  0.8&300& $\underset{(0.0261)}{-0.0062}$ & $\underset{(0.0069)}{0.0202}$ & $\underset{(0.0361)}{0.0836}$ && $\underset{(0.0330)}{-0.0094}$ & $\underset{(0.0228)}{0.1173}$ & $\underset{(0.0332)}{0.1504}$ \\
  &500& $\underset{(0.0202)}{-0.0027}$ & $\underset{(0.0028)}{0.0087}$ & $\underset{(0.0152)}{0.0405}$ && $\underset{(0.0245)}{-0.0057}$ & $\underset{(0.0110)}{0.0689}$ & $\underset{(0.0161)}{0.0876}$ \\
  &900& $\underset{(0.0149)}{-0.0024}$ & $\underset{(0.0011)}{0.0033}$ & $\underset{(0.0058)}{0.0190}$ && $\underset{(0.0189)}{-0.0040}$ & $\underset{(0.0044)}{0.0382}$ & $\underset{(0.0060)}{0.0479}$ \\
\hline
    \end{tabular}
  \end{center}
\end{table}



Examination of Table \ref{Tab 1} leads to the following conclusions.
\begin{itemize}
\item[]1) When $\rho=0$, $\widehat{\rho}$ is very small (almost zero), the estimators of $\beta(t)$ based on the proposed method and the FPCA-based method for FLM perform equally well. This result is consistent with our expectations, as the SFLM reduces to the classical FLM when $\rho=0$. 
\item[]2) When $\rho\neq0$, our proposed method produces better results than the FPCA-based method, also consistent with our expectations. The MSE of the SFLM is always smaller than the MSE of the FLM when the other settings are identical; moreover, as $\rho$ increases, the difference in the MSE between the SFLM and the FLM also increases.
\item[]3) Regardless of the value of $\rho$, the MSE of $\beta(t)$ obtained using the SFLM decreases as the sample size increases. The standard deviation of $\hat{\rho}$ also displays a decreasing pattern. Moreover, the bias of $\hat{\rho}$ is small in all cases. Similar to the numerical results presented in \cite{Lee2004Asymptotic}, the bias of $\rho$ is negative at all settings.
\item[]4) As the case in \cite{Hall2007Methodology}, $\beta(t)$ is more accurately estimated given $\gamma=1.1$ than $\gamma=2$ when the other simulation parameters are held equal. The performance of the estimator $\hat{\rho}$ is also influenced by $\alpha$. In the case in which $\gamma=1.1$, the standard deviation of $\hat{\rho}$ is smaller than that of $\gamma=2$.
\end{itemize}

\vspace{3mm}
\begin{figure}[!h]
  \begin{center}
  \subfigure{
  \label{fig2-1}
    \resizebox{4cm}{3.5cm}{\includegraphics{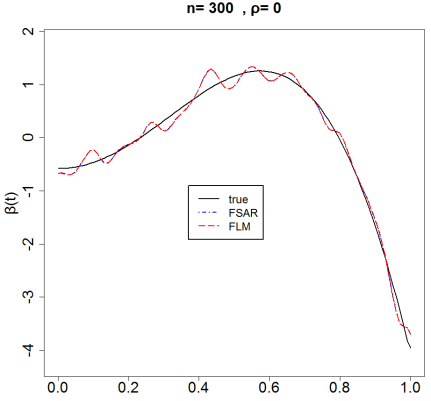}}}
    \subfigure{
    \label{fig2-2}
    \resizebox{4cm}{3.5cm}{\includegraphics{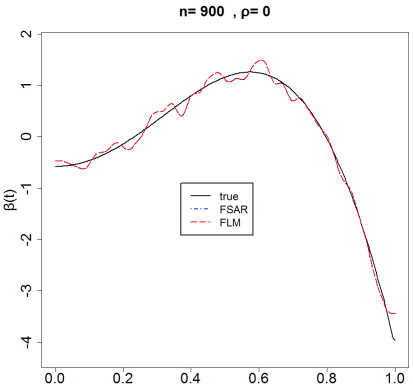}}}
    \subfigure{
    \label{fig2-3}
    \resizebox{4cm}{3.5cm}{\includegraphics{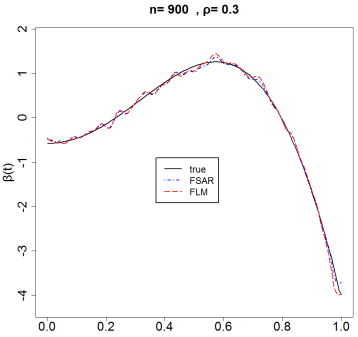}}}
      \subfigure{
  \label{fig2-1}
    \resizebox{4cm}{3.5cm}{\includegraphics{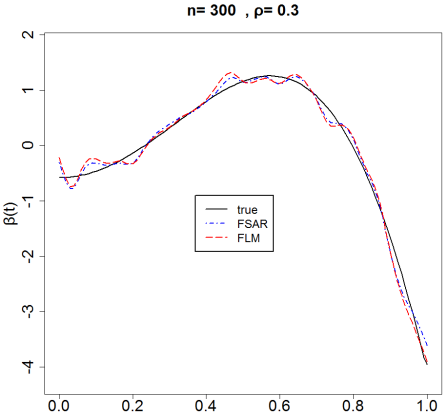}}}
    \subfigure{
    \label{fig2-2}
    \resizebox{4cm}{3.5cm}{\includegraphics{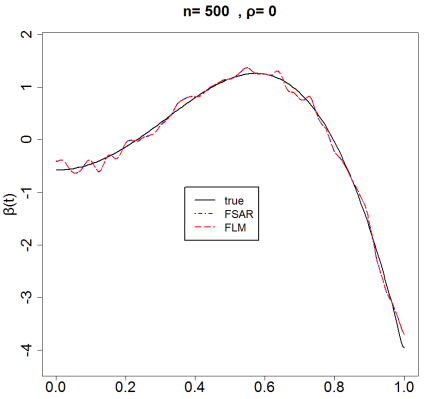}}}
    \subfigure{
    \label{fig2-3}
    \resizebox{4cm}{3.5cm}{\includegraphics{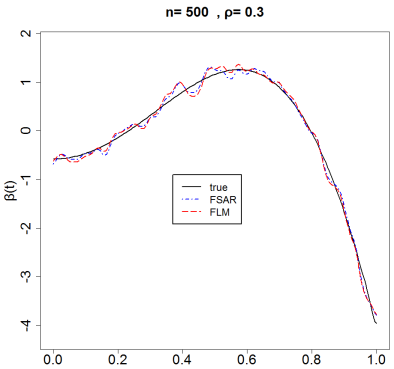}}}
    \subfigure{
  \label{fig2-1}
    \resizebox{4cm}{3.5cm}{\includegraphics{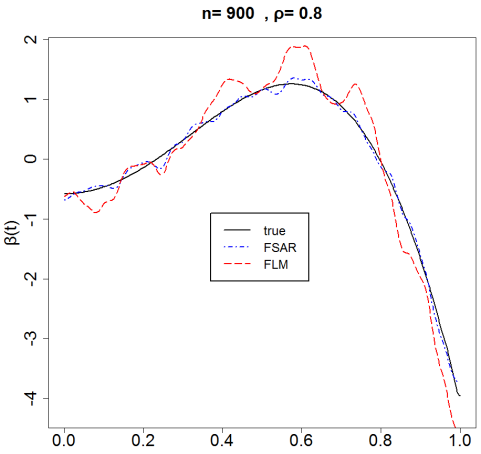}}}
    \subfigure{
    \label{fig2-2}
    \resizebox{4cm}{3.5cm}{\includegraphics{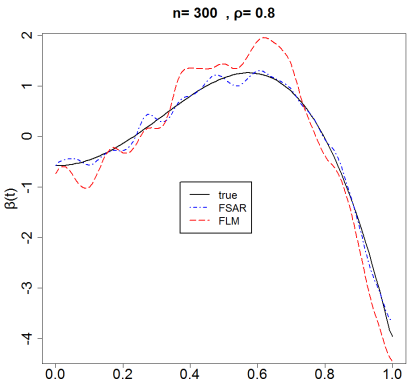}}}
    \subfigure{
    \label{fig2-3}
    \resizebox{4cm}{3.5cm}{\includegraphics{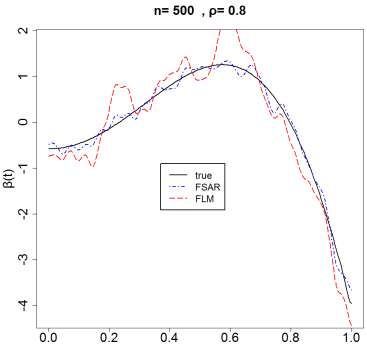}}}
  \end{center}
  \caption{Estimators of $\beta (t)$ vs the true $\beta (t)$ when the sample size $n=300,500, 900$ and $\rho=0, 0.5, 0.8$, respectively. } \label{Fig.2}
\end{figure}

Moreover, to illustrate the performance of the estimator of $\beta(t)$ intuitively, we randomly select one result from 500 repetitions, and display the estimator $\hat\beta(t)$ vs. the true value of $\beta(t)$ in Figure \ref{Fig.2} for both the SFLM and the FLM. Here, the sample size $n=300,500, 900$ and $\rho=0, 0.5, 0.8$, respectively, with $\gamma=1.1$. From Figure \ref{Fig.2}, we can obtain similar conclusions for $\beta(t)$ as those derived from Table \ref{Tab 1}.


In summary, our proposed model and estimators perform as well as the classical FLM when no spatial autocorrelation is present; on the other hand, our proposed methods outperforms the classical methods when spatial autocorrelation is present, and the difference increases with the degree of spatial autocorrelation becoming stronger. Given these results, our proposed model and estimation procedure provide a competitive alternative for the existing methods in FDA.

\section{REAL DATA ANALYSIS}\label{sec5}

\noindent In this section, we revisit the weather data presented in Section \ref{sec1} to assess the application of the SFLM. Specifically, we add a record that corresponds to the weather data for $2008$ derived from the China Meteorological Yearbook. Let the response $y_{_{i}}$ and the predictor $x_{i}(t)$ be the logarithm of the mean annual precipitation and the mean monthly temperature curve for the $i$th city between $2005$ and $2007$. We build the SFLM model as
\beq\label{realdata-sflm}
y_{i}=\rho \sum_{i\neq i^{\prime}}w_{ii^{\prime}}y_{i^{\prime}}+\int _{0}^{1}x_{i}(t)\beta(t)dt+\epsilon_{i},
\eeq
where $w_{ii^{\prime}}$ is the weight between city $i$ and $i^{\prime}$. We also build the FLM as
\beq\label{realdata-flm}
y_{i}=\int _{0}^{1}x_{i}(t)\beta(t)dt+\epsilon_{i}
\eeq
to enable comparison with the SFLM.

\begin{figure}[!h]
  \begin{center}
  \label{fig3-1}
  \resizebox{9cm}{7cm}{\includegraphics{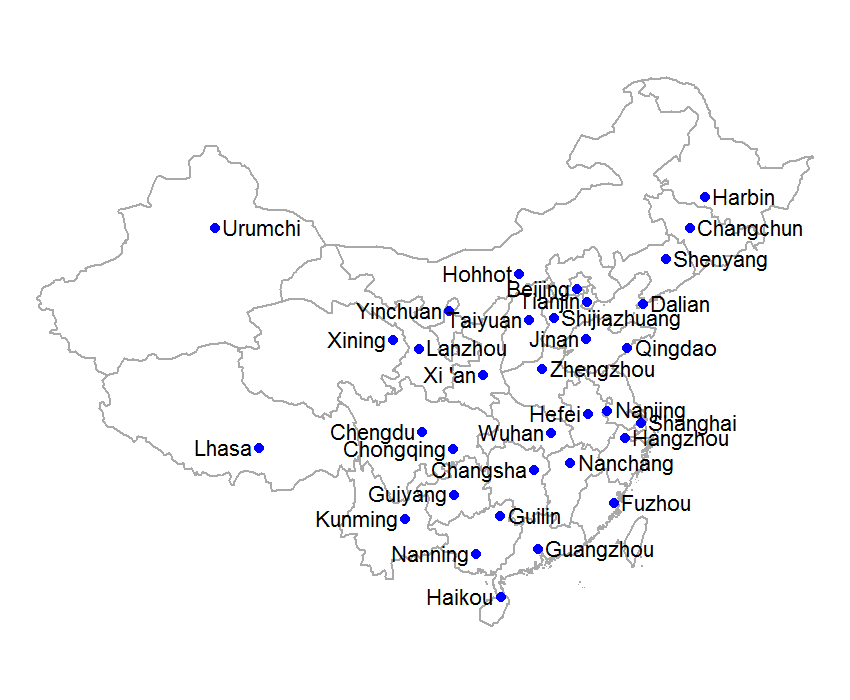}}
  \end{center}
  \vspace{-1cm}
  \caption{The locations of 34 major cities in China} \label{Fig.3}
\end{figure}

During preprocessing, we smooth the mean monthly temperatures over $3$ years using the Epanechnikov Kernel. Note that the spatial correlation between temperature curves is beyond the scope of our article. Thus, we simply presume that these functional data are independent. Moreover, the spatial weight matrix is formed using the nearest $k$ neighbours; each neighbour's weight is equal to the reciprocal of the Euclidean distance $d(i,i^\prime)$ between cities $i$ and $i^\prime$; i.e. $w_{ii^{\prime}}=d(i,i^\prime)^{-1}$. Because $k$ can take many different values, we set $k=\{2,3,4,5,6,7,8,9\}$ to enable the selection of the most appropriate number.  Moreover, we assume that, if the distance between two cities exceeds $15$, then these two cities are not neighbours of each other; i.e. $w_{ii^{\prime}}=0$. Figure \ref{Fig.3} presents the locations of 34 major cities in China. Because Urumchi is located far from the other cities, we remove its record from the weather data. The matrix is row-normalized after construction.

\begin{figure}[!h]
  \begin{center}
    \subfigure{
  \label{fig3-1}
  \resizebox{5cm}{4.7cm}{\includegraphics{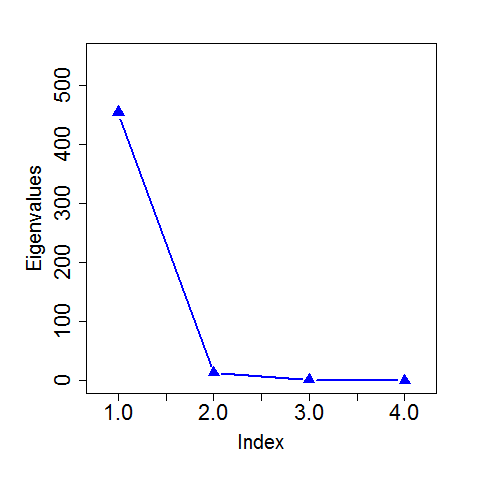}}}
  \subfigure{
  \label{fig3-2}
    \resizebox{5cm}{4.7cm}{\includegraphics{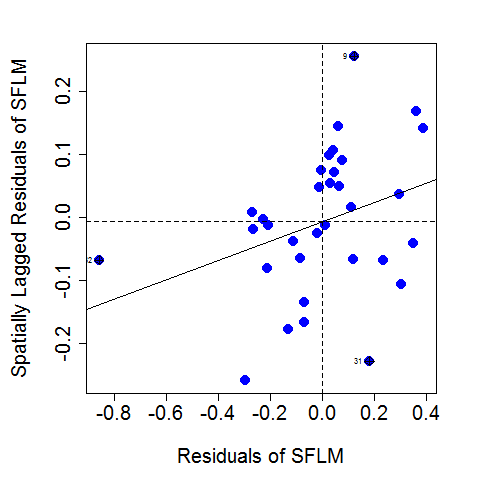}}}
    \subfigure{
    \label{fig3-3}
    \resizebox{5cm}{4.7cm}{\includegraphics{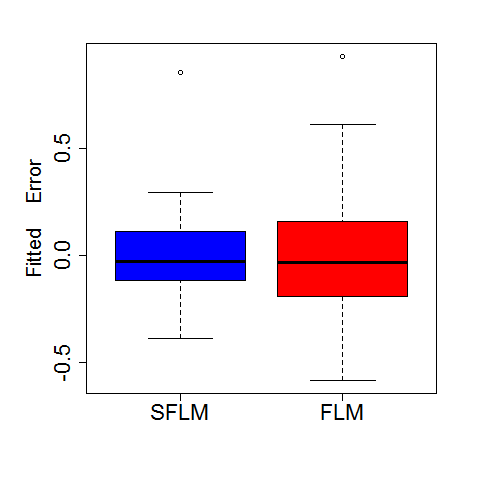}}}
    \subfigure{
    \label{fig3-4}
    \resizebox{5cm}{4.7cm}{\includegraphics{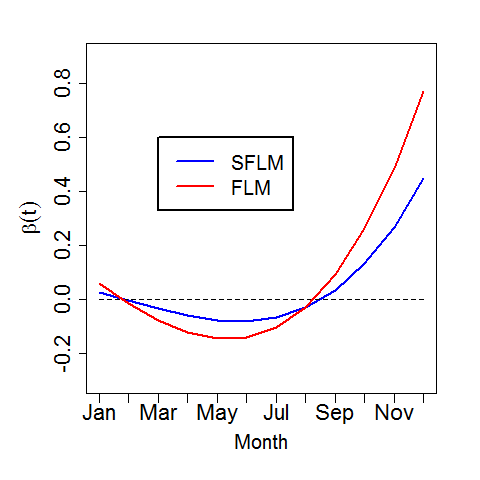}}}
  \end{center}
  \vspace{-3mm}
  \caption{Eigenvalues of sample covariance function (top left), Moran I scatter plot of residuals of SFLM (top right), fitted error of SFLM and FLM (bottom left), and estimated $\beta(t)$ of SFLM and FLM (bottom right), respectively.} \label{Fig.4}
\end{figure}

Figure \ref{Fig.4} (top-left) shows the eigenvalues of the sample covariance function. The eigenvalues clearly decay quickly, and the first two eigenvalues account for $99\%$ of the total variance. Therefore, we consider two candidate truncation parameters, $1$ and $2$. In the first case, only one covariate is contained in the truncated SFLM; in the second case, two covariates are included. Combining the $8$ possible weight matrices mentioned above with these two parameters, we have $16$ candidates for the truncated SFLM. Because 16 is relatively small, we compute their posterior model probabilities directly, instead of using a strategic stochastic Markov chain process to search for an appropriate model. The log-likelihood function values and Bayesian model probability ratios of each model are displayed in Table \ref{Tab_3}.


\begin{table}[!h]
\footnotesize
\caption{ The values of the log-likelihood function and the Bayesian model probability ratios of the $16$ candidates of the truncated SFLM, denoted $L$ and $B$, respectively. Specifically, the Bayesian model probability ratio is obtained by dividing the posterior probability of the current model by the posterior probability of the model for which $m$ and $k$ are $1$ and $2$.}
  \label{Tab_3}
  \setlength\tabcolsep{6pt}
  \renewcommand{\arraystretch}{1.5}
\bc
\begin{tabular}[H]{ccccccccccccccccccc}
\specialrule{0.05em}{1pt}{1pt}
$\bm{m/k}$ &  & $\bm{k=2}$ & $\bm{k=3}$ & $\bm{k=4}$ & $\bm{k=5}$ & $\bm{k=6}$ & $\bm{k=7}$ & $\bm{k=8}$ & $\bm{k=9}$ \\
\specialrule{0.05em}{1pt}{1pt}
$\bm{m=1}$ & $L$ & -9.38 & -8.54 & -5.42 & -3.72 & -4.13 & -4.10 & -4.47 & -4.94 &\\ \specialrule{0.0em}{1pt}{1pt}
& $B$ & 1 & 2.46 & 44.37 & 198.31 & 141.03 & 145.42 & 108.51 & 69.89 &\\ \specialrule{0.05em}{1pt}{1pt}
$\bm{m=2}$ & $L$ & -4.42 & -3.28 & -1.51 & -1.11 & -1.90 & -2.15 & -2.25 & -2.50 & \\ \specialrule{0.0em}{1pt}{1pt}
& $B$ & 90.39 & 302.32 & 1755.32  & 2253.14 & 1163.4 & 943.31 & 887.44 & 695.95 &\\ \specialrule{0.05em}{1pt}{1pt}
\end{tabular}\ec
\end{table}

\begin{table}[!h]
\footnotesize
\caption{ The fitting and prediction results of the SFLM and the FLM. Here, the fitting error and prediction error are evaluated separately in terms of the mean square error of the fitted values and the prediction values.}
  \label{Tab_2}
  \setlength\tabcolsep{6pt}
  \renewcommand{\arraystretch}{1.5}
\bc
\begin{tabular}[H]{ccccccc}
\specialrule{0.05em}{1pt}{1pt}
  & models& $\hat{\rho}$ &  Moran's I statistic(residuals) & MSE(fitted error)& MSE(prediction error)
  \\
  \specialrule{0.05em}{1pt}{1pt}
  & FLM& ---& 0.41  &  0.33 & 0.12\\ \specialrule{0em}{-3.5pt}{-3.5pt}
  & SFLM &  0.58   & 0.15 & 0.24 & 0.10\\ \specialrule{0em}{-3.5pt}{-3.5pt}
  \specialrule{0.05em}{7pt}{7pt}
\end{tabular}\ec
\end{table}
The results presented in Table \ref{Tab_3} can be summarized as follows. The matrix with $5$ nearest neighbours is the most appropriate for our model, and when $m=2$, the SFLM yields a greater posterior probability than when $m=1$. We display the fitting results of the SFLM and the FLM with $m=2,k=5$ in Table \ref{Tab_2} and Figure \ref{Fig.4}. To assess the predictive ability of the SFLM, we apply the fitting result to the temperature observations made in $2008$ to determine the annual precipitation in the same year. The fitting results and prediction errors are displayed in Table \ref{Tab_2} and Figure \ref{Fig.4}.


\setlength{\textfloatsep}{1\normalbaselineskip}

Recall from Section \ref{sec1} that Figure \ref{Fig.1} reflects significant spatial autocorrelation among the annual precipitation values for the different cities, and the FLM can not efficiently address this correlation. The Moran's I scatterplot of residuals of the SFLM in Figure \ref{Fig.4} suggests that a majority of the spatial autocorrelation in responses has been removed. We can also conclude from Figure \ref{Fig.4} that the SFLM has reduced the fitting error substantially when compared to the FLM.

As for the estimated parameters, $\hat{\rho}$ is $0.58$, and the corresponding P value is smaller than $0.001$; the slope functions of the two models are provided in Figure \ref{Fig.4} (to the right). The two curves of the estimated $\beta(t)$ have similar shapes, and the $\hat{\beta}(t)$ of the SFLM is smoother than that of the FLM. We conclude that precipitation is much more strongly influenced by temperature during the winter than in the other seasons; moreover, under SFLM, the precipitation received by each city is less affected by temperature over the year as a whole.

To illustrate the superiority of the SFLM relative to the FLM, we also compare their predictive abilities. We use the data from $2005$ to $2007$ to predict the precipitation in $2008$. The MSE values of the predicted $y_{i}$ under the SFLM and FLM are $0.10$ and $0.12$, respectively, which reflects substantial improvement. 

\section{CONCLUSION AND DISCUSSION}\label{sec6}

\noindent The FLM is popular in studies of links between a scalar response and functional predictors. However, the existing FLM cannot address the dependencies that arise due to the presence of a network structure. We propose a powerful spatial functional linear model that integrates the advantages of FLM in handling high-dimensional data and SAR model in coping with spatial dependencies. A simple estimation method is developed to obtain the estimators of the spatial autoregressive parameter and the slope function. Our simulation study demonstrates the consistency of the proposed estimators. In particular, the new estimators perform as well as the FPCA-based estimator of the FLM when the spatial autoregressive parameter equals zero; on the other hand, the new methods outperform the existing ones when spatial autocorrelation is present. An examination of a real dataset demonstrates the superiority of the SFLM over the FLM. From this perspective, our proposed model and estimation procedure represent competitive alternatives to the FLM.

Note that network structure is indeed what is considered in the SFLM, and the concept of network-structured data is more general than the spatially correlated data considered in this article. For example, social network data display network structure, but they are not spatially correlated. In fact, the weather data studied here are spatially dependent in terms of their network structure, which provides a new perspective on spatial functional data. Moreover, we discuss the SFLM under the assumption that only one functional predictor is involved. However, the SFLM with multiple functional predictors also deserve attention. Based on the estimation method introduced in Section \ref{sec3}, the methods presented here can easily be generalized to the SFLM with multiple functional predictors. Functional variable selection can then be conducted. These steps are beyond the scope of the current paper and will be investigated further in subsequent work.
\section*{Acknowledgements}

\noindent This research was financially supported by the National Natural Science Foundation
of China under grant nos. 71420107025 and 11701023.

\bibliography{SFLM-ref}

\begin{thebibliography}{25}
\providecommand{\natexlab}[1]{#1}
\providecommand{\url}[1]{\texttt{#1}}
\expandafter\ifx\csname urlstyle\endcsname\relax
  \providecommand{\doi}[1]{doi: #1}\else
  \providecommand{\doi}{doi: \begingroup \urlstyle{rm}\Url}\fi

\bibitem[Aguilera-Morillo et~al.(2016)Aguilera-Morillo, Durb芍n, and
  Aguilera]{Aguilera2016Prediction}
{\rm Aguilera-Morillo, M.~C., Durb芍n, M., {\rm and} Aguilera, A.~M.} (2016).
\newblock Prediction of functional data with spatial dependence: a penalized
  approach.
\newblock \emph{Stochastic Environmental Research \& Risk Assessment}, pages
  1--16.

\bibitem[Aneiros-P\'{e}rez and Vieu(2006)]{Aneiros2006Semi}
{\rm Aneiros-P\'{e}rez, G. {\rm and} Vieu, P.} (2006).
\newblock Semi-functional partial linear regression.
\newblock \emph{Statistics and Probability Letters}, {\bf 76}\penalty0 (11),
  \penalty0 1102--1110.

\bibitem[Brockhaus et~al.(2015)Brockhaus, Scheipl, Hothorn, and
  Greven]{brockhaus2015functional}
{\rm Brockhaus, S., Scheipl, F., Hothorn, T., {\rm and} Greven, S.} (2015).
\newblock The functional linear array model.
\newblock \emph{Statistical Modelling}, {\bf 15}\penalty0 (3), \penalty0
  279--300.

\bibitem[Cai and Hall(2006)]{TT2006Prediction}
{\rm Cai, T. {\rm and} Hall, P.} (2006).
\newblock Prediction in functional linear regression.
\newblock \emph{The Annals of Statistics}, {\bf 34}\penalty0 (5), \penalty0
  2159--2179.

\bibitem[Case(1991)]{Case1991Spatial}
{\rm Case, A.~C.} (1991).
\newblock Spatial patterns in household demand.
\newblock \emph{Econometrica}, {\bf 59}\penalty0 (4), \penalty0 953--965.

\bibitem[Escabias et~al.(2004)Escabias, Aguilera, and
  Valderrama]{Escabias2004Principal}
{\rm Escabias, M., Aguilera, A.~M., {\rm and} Valderrama, M.~J.} (2004).
\newblock Principal component estimation of functional logistic regression:
  discussion of two different approaches.
\newblock \emph{Journal of Nonparametric Statistics}, {\bf 119}\penalty0 (1),
  \penalty0 1--6.

\bibitem[Ferraty et~al.(2013)Ferraty, Goia, Salinelli, and
  Vieu]{ferraty:hal-00966065}
{\rm Ferraty, F., Goia, A., Salinelli, E., {\rm and} Vieu, P.} (2013).
\newblock Functional projection pursuit regression.
\newblock \emph{test}, {\bf 22}\penalty0 (2), \penalty0 293--320.

\bibitem[Hall and Horowitz(2007)]{Hall2007Methodology}
{\rm Hall, P. {\rm and} Horowitz, J.~L.} (2007).
\newblock Methodology and convergence rates for functional linear regression.
\newblock \emph{Annals of Statistics}, {\bf 35}\penalty0 (1), \penalty0 70--91.

\bibitem[Horv{\'a}th and Kokoszka(2012)]{horvath2012inference}
{\rm Horv{\'a}th, L. {\rm and} Kokoszka, P.} (2012).
\newblock \emph{Inference for functional data with applications}, volume 200.
\newblock Springer Science \& Business Media.

\bibitem[James(2002)]{RSSB:RSSB342}
{\rm James, G.~M.} (2002).
\newblock Generalized linear models with functional predictors.
\newblock \emph{Journal of the Royal Statistical Society: Series B (Statistical
  Methodology)}, {\bf 64}\penalty0 (3), \penalty0 411--432.

\bibitem[Kelejian and Prucha(1999)]{Kelejian1999A}
{\rm Kelejian, H.~H. {\rm and} Prucha, I.~R.} (1999).
\newblock A generalized moments estimator for the autoregressive parameter in a
  spatial model.
\newblock \emph{International Economic Review}, {\bf 40}\penalty0 (2),
  \penalty0 509每533.

\bibitem[Lee(2004)]{Lee2004Asymptotic}
{\rm Lee, L.~F.} (2004).
\newblock Asymptotic distributions of quasi-maximum likelihood estimators for
  spatial autoregressive models.
\newblock \emph{Econometrica}, {\bf 72}\penalty0 (6), \penalty0 1899--1925.

\bibitem[Lee(2007)]{Lee2007GMM}
{\rm Lee, L.~F.} (2007).
\newblock Gmm and 2sls estimation of mixed regressive, spatial autoregressive
  models.
\newblock \emph{Journal of Econometrics}, {\bf 137}\penalty0 (2), \penalty0
  489--514.

\bibitem[Lesage and Pace(2009)]{Lesage2009Introduction}
{\rm Lesage, J. {\rm and} Pace, R.~K.} (2009).
\newblock \emph{Introduction to Spatial Econometrics}.
\newblock CRC Press.

\bibitem[Lesage and Parent(2007)]{Lesage2007Bayesian}
{\rm Lesage, J.~P. {\rm and} Parent, O.} (2007).
\newblock Bayesian model averaging for spatial econometric models.
\newblock \emph{Geographical Analysis}, {\bf 39}\penalty0 (3), \penalty0
  241每267.

\bibitem[Liu et~al.(2017)Liu, Wang, and Cao]{Liu2017Estimating}
{\rm Liu, B., Wang, L., {\rm and} Cao, J.} (2017).
\newblock Estimating functional linear mixed-effects regression models.
\newblock \emph{Computational Statistics and Data Analysis}, {\bf 106},
  \penalty0 153--164.

\bibitem[Olubusoye et~al.(2016)Olubusoye, Korter, and
  Salisu]{Olubusoye2016MODELLING}
{\rm Olubusoye, O.~E., Korter, G.~O., {\rm and} Salisu, A.~A.} (2016).
\newblock Modelling road traffic crashes using spatial autoregressive model
  with additional endogenous variable.
\newblock \emph{Statistics in Transition New}, {\bf 17}, \penalty0 659--670.

\bibitem[Ord(1975)]{Ord1975Estimation}
{\rm Ord, K.} (1975).
\newblock Estimation methods for models of spatial interaction.
\newblock \emph{Journal of the American Statistical Association}, {\bf
  70}\penalty0 (349), \penalty0 120--126.

\bibitem[Paganoni and Sangalli(2017)]{paganoni2017functional}
{\rm Paganoni, A.~M. {\rm and} Sangalli, L.~M.} (2017).
\newblock Functional regression models: Some directions of future research.
\newblock \emph{Statistical Modelling}, {\bf 17}\penalty0 (1-2), \penalty0
  94--99.

\bibitem[Pinedar赤os and Giraldo(2016)]{Pinedar2016Functional}
{\rm Pinedar赤os, W. {\rm and} Giraldo, R.} (2016).
\newblock Functional sar model.

\bibitem[Qu and Lee(2015)]{Qu2015Estimating}
{\rm Qu, X. {\rm and} Lee, L.~F.} (2015).
\newblock Estimating a spatial autoregressive model with an endogenous spatial
  weight matrix.
\newblock \emph{Journal of Econometrics}, {\bf 184}\penalty0 (2), \penalty0
  209--232.

\bibitem[Ram車n et~al.(2017)Ram車n, Pedro, and Jorge]{Giraldo2017Spatial}
{\rm Ram車n, G., Pedro, D., {\rm and} Jorge, M.} (2017).
\newblock Spatial prediction of a scalar variable based on data of a functional
  random field.
\newblock \emph{Comunicaciones en Estad赤stica}, {\bf 10}\penalty0 (2),
  \penalty0 315--344.

\bibitem[Ramsay and Silverman(2002)]{RamsaySilverman-506}
{\rm Ramsay, J.~O. {\rm and} Silverman, B.~W.} (2002).
\newblock \emph{Applied functional data analysis: methods and case studies},
  volume~77.
\newblock Springer.

\bibitem[Topa(2001)]{Topa2001Social}
{\rm Topa, G.} (2001).
\newblock Social interactions, local spillovers and unemployment.
\newblock \emph{Review of Economic Studies}, {\bf 68}\penalty0 (2), \penalty0
  261--295.

\bibitem[Zhou et~al.(2013)Zhou, Wang, and Wang]{Zhou2013FUNCTIONAL}
{\rm Zhou, J., Wang, N.~Y., {\rm and} Wang, N.} (2013).
\newblock Functional linear model with zero-value coefficient function at
  sub-regions.
\newblock \emph{Statistica Sinica}, {\bf 23}\penalty0 (1), \penalty0 25--50.

\end{thebibliography}

\makeatletter
\addtolength{\@fpsep}{-12pt}
\makeatother

\makeatletter
\addtolength{\@fpsep}{-10pt}
\makeatother

\vspace{-10mm}

\end{document}